\documentstyle[aps]{revtex}
\begin{document}
\newtheorem{th}{Theorem}
\newtheorem{lem}[th]{Lemma}

\title{
Elements of Nonlinear Quantum Mechanics (II):\\
Triple bracket generalization of quantum mechanics
}
\author{Marek Czachor~\cite{*}}
\address{
Pracownia Dielektryk\'{o}w i P\'{o}\l przewodnik\'{o}w
Organicznych\\
Wydzia{\l}  Fizyki Technicznej i Matematyki Stosowanej\\
 Politechnika Gda\'{n}ska\\
ul. Narutowicza 11/12, 08-952 Gda\'{n}sk, Poland
}
\maketitle
\begin{abstract}
An extension of quantum mechanics to a generalized Nambu
dynamics leads to a new version of nonlinear quantum mechanics.
The time evolution of states is given in here by a
triple bracket generalization of the Liouville-von Neumann
equation, where one of the generators is an average energy, and
the other is a measure of entropy. A nonlinear evolution can
occur only for mixed states, and for systems that are described
by R\'enyi $\alpha$-entropies with $\alpha\neq 2$. The case
$\alpha=2$ corresponds to ordinary, linear quantum mechanics.
Since $\alpha=2$ entropy is the only entropy characterizing
systems which cannot gain information, the nonlinear dynamics
corresponds to ``observers", that is,
 systems that {\em can\/} gain
information. The new formulation of nonlinear quantum mechanics
is free from difficulties found in earlier attempts.
The connection of linearity with possibilities of gaining
information is in a striking agreement with the ideas of
Wigner formulated in his paradox of a
friend.
\end{abstract}

\pacs{03.65Bz, Ca}
\section{Introduction}

In the first part of this article \cite{I} I have described the
fundamental  theoretical difficulties of nonlinear
quantum mechanics (NLQM) based on a nonlinear Schr\"odinger
equation. In the present paper I will present a different
generalization of quantum mechanics (QM) where the nonlinear
evolution never occurs for {\em pure\/} states (the
Schr\"odinger equation is hence always linear in this framework)
but, instead, may appear, under some circumstances, for {\em
mixed\/} states. We will see that
the new approach will be free from
the difficulties discussed in Ref.\cite{I}.

The proposed generalization is based on the idea of rewriting
the Liouville-von Neumann equation in a triple bracket form,
introduced by Bia\l ynicki-Birula and
Morrison \cite{Mor}. The triple bracket is an infinite
dimensional analog of the
Nambu bracket \cite{Nambu} where, as opposed to the structure
constants $\epsilon_{klm}$ of the rotation algebra appearing in
the original Nambu bracket, the
structure constants correspond to
 some infinite-dimensional Lie algebra. In
the original Nambu paper an evolution of a physical system
(a rigid rotator) is
generated by two ``Hamiltonian functions", the energy $H$ and $J$, where
the latter is the Casimir of $so(3)$ (squared angular momentum).
The metric tensor used for constructing the Casimir is, as
usual,  the one
related to the Killing form \cite{BR}.
In the triple bracket formulation of QM the analog of $J$ is
the Casimir $S=1/2{\rm Tr}(\rho^2)$ which also can be written as
$g^{ab}\rho_a\rho_b$ although, as we shall see later, the metric
$g^{ab}$ is no longer given by the Killing-Cartan tensor (which does
not exist in this case). The Casimir
$S$ was termed in Ref.\cite{Mor} the {\em entropy\/}. It will be
argued below that the assignment of the name ``entropy" to $S$
should not be regarded as accidental, but as a reflection of a
deeper principle relating dynamics with information.

The fact
that some sort of such a relationship should be present in QM
follows already from the Copenhagen interpretation of
a measurement (reduction of a state vector), but my approach
will be essentially different and closer in spirit to Wigner's
paradox of a friend \cite{MB}.  Let me recall that
Wigner, in order to solve the
paradox, concluded that a conscious observation must be
accompanied by a {\it nonlinear\/} evolution in the space of the
observer's states. Even though the argumentation of Wigner looks
convincing, it seems that standard quantum theories do not leave
room for a physical principle of that kind. It is surprising
that the triple bracket formalism {\em does\/}
 lead quite naturally to this
phenomenon, if we seriously treat the intuitions of Bia\l
ynicki-Birula and Morrison that $S$ {\em is\/} a measure of
quantum entropy.

Putting things more modestly, one can say that the results of this
paper, even if their interpretation will
turn out inadequate,
show that the
structure of quantum dynamics may be a part of a more general,
nonlinear framework.

The structure of the paper is the following. In Sec.~II I discuss
various measures of information (entropies) and their possible
relationship
with quantum mechanics. With this background we will be able to
understand why  ${\rm Tr}(\rho^2)$ is a natural measure of information
characterizing systems that {\em cannot gain information\/} (and,
accordingly, for
which the paradox of a friend cannot be formulated). In Sec.~III, I
present a triple bracket  formulation of QM and introduce the
composite index form of the equation of motion.
In Sec.~IV I will discuss an extension of the formalism to nonlinear
theories in which the second ``Hamiltonian function", here interpreted
as a measure of entropy, is represented by Casimirs of order higher
than 2. Such Casimirs correspond quite naturally to higher order
$\alpha$-entropies discussed in Sec.~II. The main technical results
of this section are  theorems on non-existence of ``faster-than-light
telegraphs" and on conservation of positivity of $\rho_t$ by the
nonlinear evolution equation resulting from the triple bracket
formalism. It is also shown that homogeneity preserving generalizations
of $\alpha$-entropies lead to {\em linear\/} evolution of {\em
pure\/} states, and that the $\alpha\to 1$ limit of $\alpha$-entropies
(the Shannon limit) can be regarded as a kind of classical limit
for the generalized QM. Finally, I discuss various possibilities
of describing composite systems that consist of subsystems
described by different entropies. The paper is concluded with
a remark on the complementarity principle in nonlinear QM and
an explicitly relativistic formulation is given in the Appendix.

\section{Measures of information and quantum mechanics}

A {\it logarithmic measure \/} of information was introduced by
R.~V.~Hartley in 1928 \cite{Hartley}. According to him, to
characterize an element of a set of size $N$ we need $\log_2N$
units of information. It follows that a unit of information (1 bit) is
the amount of information necessary for a characterization of a
pair. Of course, one can choose also other units such that the
unit is the amount of information necessary for a
characterization of a set with $0<k\in {\bf N}$ elements, or even
with $0<r\in {\bf R}$ elements in average. The respective
measures of information in arbitrary units $a$ are
$\log_aN$. The most important feature  of the logarithmic
information measure is its additivity: If a set $E$ is a
disjoint union of $M$ $N$-tuples $E_1,\dots,E_M$, then we can
specify an element of this $MN$-element set $E$ in two steps:
First we need $\log_aM$ units of information to describe which
$E_k$ of the sets $E_1,\dots,E_M$ contains the element, then we
need $\log_aN$ further units to tell which element of this $E_k$
is the considered one. The information necessary for a
characterization of an element of $E$ is the sum of the partial
informations: $\log_aMN=\log_aM+\log_aN$. Next step in the
developement of the measures of information was done
independently by C.~E.~Shannon \cite{Shannon} and N.~Wiener
\cite{Wiener} in 1948 who derived a formula analogous to
Boltzman's entropy. Their formula has the following heuristic
motivation. Let $E$ be the disjoint union of the sets
$E_1,\dots, E_n$ having $N_1,\dots,N_n$ elements respectively
$\bigl(\sum_{k=1}^{n} N_k=N\bigr)$. Let us suppose that we are
interested only in knowing the subset $E_k$. (This is typical
for classical statistical problems in physics: Statistical quantities
depend on classes of microscopic conditions and not on single
microscopic properties.) The information characterizing an
element of $E$ consists of two parts: The first specifies the
subset $E_k$ containing this particular element and the second
locates it within $E_k$. The amount of the second piece of
information is, by Hartley formula, $\log_aN_k$ thus depends on
the index $k$. On the other hand, to specify an element of $E$
we need $\log_aN$ units of information. The amount necessary for
the specification of the set $E_k$ is therefore
\begin{equation}
I_k=\log_aN -
\log_aN_k=\log_a\frac{N}{N_k}=\log_a\frac{1}{p_k}.
\end{equation}
It follows that the amount of information {\it received by
learning\/} that a single
 event of probability $p$ took place equals
\begin{equation}
I(p)=\log_a\frac{1}{p}.
\end{equation}
In statistical situations  measured quantities correspond
to averages of random variables. Therefore the average
information is
\begin{equation}
I=\sum_k p_k\log_a\frac{1}{p_k}.\label{Shannon}
\end{equation}
This is the Shannon's formula and $I$ is called the {\it
entropy\/} of the probability distribution $\{p_1,\dots,p_n\}$.
 If all the probabilities are
equal $1/N$ then the Shannon's formula is equal to the Hartley's
one. The mean we have applied is the so-called linear mean.
R\'enyi observed that there exist information theoretic problems
where the measures of information are those obtained by more
general ways of averaging --- the Kolmogorov--Nagumo function
approach \cite{Renyi}.
Let $\varphi$ be a monotonic function  on real numbers.
The Kolmogorov--Nagumo average information can be defined by
means of
$\varphi$ as
\begin{equation}
I=\varphi^{-1}\Biggl(\sum_k
p_k\varphi\Bigl(\log_a\frac{1}{p_k}\Bigr)\Biggr).
\end{equation}
If the generalized information measure is to satisfy the
postulate of additivity, $\varphi$
must be
a linear or exponential
function. The linear function corresponds to Shannon's
information. The exponential functions provide a large class of
new measures of information. Consider a function
$\varphi(x)=a^{(1-\alpha)x}$. We can always choose the units of
information in such a way that
\begin{equation}
I=\varphi^{-1}\Biggl(\sum_k
p_k\varphi\Bigl(\log_a\frac{1}{p_k}\Bigr)\Biggr)=
\frac{1}{1-\alpha}\log_a\Bigl(\sum_kp_k^\alpha\Bigr)=
\log_a\Biggl(\Bigl(\sum_kp_k^\alpha\Bigr)^{1/(1-\alpha)}\Biggr).
\label{alpha}
\end{equation}
For $p_k=1/N$ we obtain again the Hartley formula. Formula
(\ref{alpha}) describes R\'enyi's $\alpha$-entropy which, from
now on, will be denoted $I_\alpha({\cal P})$, where $\cal P$
denotes the probability distribution. We see that
the essential part of the definition is played by
\begin{equation}
I^*_\alpha({\cal P})=a^{I_\alpha({\cal P})}=
\Bigl(\sum_kp_k^\alpha\Bigr)^{1/(1-\alpha)}
\end{equation}
which is independent of the choice of the unit $a$.
To
distinguish between $\alpha$-entropy and $I^*_\alpha({\cal P})$ we
shall call the latter $\alpha^*$-entropy ($*$ will remind us
that this quantity is multiplicative in opposition to the additivity
of $I_\alpha({\cal P})$). (The observation that what is in fact
informationally fundamental in $I_\alpha({\cal P})$ is
$I^*_\alpha({\cal P})$ is strenghtened by Dar\'oczy's definition
of {\it entropy of order\/} $\alpha$ \cite{Dar} defined as
\begin{equation}
(2^{1-\alpha} -1)^{-1}\Bigl(\sum_kp_k^\alpha-1\Bigr).
\end{equation}
This expression possesses many ordinary properties of the
entropy and in the limit $\alpha\to 1$ becomes, the so-called
Shannon's information function.)

The limit $\alpha\to 1$ is interesting also for
$\alpha$-entropies. It can be shown that
$I_1=\lim_{\alpha\rightarrow 1}I_\alpha$ equals Shannon's
entropy.

$I_\alpha({\cal P})$ is a monotonic, decreasing function of
$\alpha$. For negative $\alpha$ $I_\alpha({\cal P})$ tends to
infinity if one of $p_k$ tends to zero. This property excludes
$\alpha<0$ because adding  a new event of probability 0 to a
probability distribution, what does not change the probability
distribution, turns $I_\alpha({\cal P})$ into infinity.

A fundamental notion in information theory is the {\it gain
of information\/}. Consider an experiment whose results are
$A_1,\dots,A_n$ having probabilities $p_k=P(A=A_k)$. We observe
an event $B$ related to the experiment and obtain a result
$B=B_l$. Now the conditional probabilities are
$p_{kl}=P(A=A_k|B=B_l)$. Consider now a system (an ``observer")
whose information is measured by some $\alpha$-entropy. How much
information about the random variable $A$ has he received by
observation of $B=B_l$? The amount of information he would have
obtained by observing $A=A_k$ would be equal to
\begin{equation}
\log_a\frac{1}{p_k}
\end{equation}
if he had not measured $B$. After having observed $B=B_l$ the
amount of information he would have obtained by observing
$A=A_k$ would be
\begin{equation}
\log_a\frac{1}{p_{kl}}.
\end{equation}
It follows that the measurement of $B=B_l$ has given him already
\begin{equation}
\log_a\frac{1}{p_k}-\log_a\frac{1}{p_{kl}}=\log_a\frac{p_k}
{p_{kl}}\label{uncert}
\end{equation}
units of information about $A$. The expression (\ref{uncert}) is
called the decrease of uncertainty  about
$A=A_k$ by observing $B=B_l$. We define the gain of information
about $A$, obtained when the probability distribution
$\{p_k\}$ is replaced by $\{p_{kl}\}$, by
\begin{equation}
\varphi^{-1}\Biggl(\sum_kp_{kl}\varphi\Bigl(\log_a\frac{p_k}
{p_{kl}}\Bigr)\Biggr)=
\frac{1}{1-\alpha}\log_a\Bigl(\sum_k
\frac{p_{kl}^{2-\alpha}}{p_k^{1-\alpha}}
\Bigr).\label{gain}
\end{equation}
If we define the increase of the uncertainty by minus decrease
of uncertainty we can calculate the average ``loss of
information" defined by
\begin{equation}
\varphi^{-1}\Biggl(\sum_kp_{kl}\varphi\Bigl(\log_a\frac
{p_{kl}}{p_k}
\Bigr)\Biggr)=
\frac{1}{1-\alpha}\log_a\Bigl(\sum_k
\frac{p_{kl}^{\alpha}}{p_k^{\alpha-1}}
\Bigr).\label{loss}
\end{equation}
For Shannon's entropy the gain is minus the loss. For
$\alpha$-entropies the two concepts are inequivalent.

The gain of information defined by (\ref{gain}) for $\alpha>2$
has the same pathological properties as $I_\alpha$ for
$\alpha<0$ so, it seems, cannot be consistently applied unless
we restrict $0\leq \alpha\leq 2$. This is
the reason why R\'enyi defined the gain of information as minus
the loss, although such a definition is less netural. From the
viewpoint of our quantum mechanical
applications the situation is not so clear, however, and the
following argument shows that $\alpha=2$ is a natural
value limiting $\alpha$-s from above.

When we speak about information,
what we have in mind is not the subjective ``information"
possessed by a particular, animate observer. In reality the
information contained in an observation is a quantity
independent of the fact whether it does or does not reach the
perception of the observer (be it a man, some registering
device, a computer, or some other physical system). On the other hand,
different kinds of entropies introduced above may be
characteristic for different systems. The entropy (information)
is objective
in the same sense as probability, and in the same sense it is
reasonable to expect that there are classical and quantum
informations, as there are classical and quantum probabilities.

The procedure leading to  the notion of the decrease of
uncertainty assumes implicitly that after each measurement of a
random variable, here $B$, one can always proceed further in
getting information about $A$, and that the procedure terminates
when we know everything about the state of the system. In
classical world this final state of knowledge means no
uncertainties. Therefore, classically, if there is some lack of
knowledge about a system, then there exists, in principle, a
possibility of {\em gaining information\/}.
Putting it more formally, we can say that an
information characterizing a {\it classical\/} system should
allow for different gains of information  in
different situations.
The quantum mechanical no-hidden-variables postulate
means that the probabilistic description of a quantum system
does not
follow from our lack of knowledge about the system.
This suggests that a quantum information,
characterizing a quantum system,
might be of such a
kind that its corresponding gain of information is zero under
all circumstances. It is tempting to develop this hypothesis a
little and find whether a measure of information possessing this
property exists.

The Shannon's information gain is given by
\begin{equation}
-\sum_kp_{kl}\log_a\frac{p_{kl}}{p_k}
\end{equation}
and vanishes only if $A$ and $B$ are independent. So this case
can be excluded because we want the gain of information to be 0 for
all probability distributions (this excludes also the von
Neumann entropy). For $\alpha$-entropies we find
that the vanishing of (\ref{gain}) implies
\begin{equation}
\sum_k
\frac{p_{kl}^{2-\alpha}}{p_k^{1-\alpha}}=
\sum_k p_k\bigl(p_{kl}^{2-\alpha}p_k^{\alpha -2}\bigr)=1
\end{equation}
which can hold for all $p_k$ and $p_{kl}$ if and only if
$\alpha=2$. It follows that the only candidate for the quantum
entropy is the R\'enyi's 2-entropy which reads
\begin{equation}
-\log_a\Bigl(\sum_k p_k^2\Bigr).
\end{equation}
Expressing the probabilities by means of a density matrix and
choosing the unit of information with $a=e$ we
obtain
\begin{equation}
I_2[\rho]=-\ln{\rm Tr}(\rho^2).
\end{equation}
This kind of entropy is sometimes considered as an alternative
to von Neumann's entropy \cite{Fano}. Our reasoning, based on
the assumption that an {\em ordinary
quantum\/} system should not have a
possibility of gaining information, selects this entropy in a
unique way. It is clear, from the perspective of the Wigner's
paradox of a friend, that {\em observers\/}, who can gain
information, should be described by $\alpha\neq 2$-entropies.

\section{Poissonian Formulation of  Quantum Mechanics}
\label{sec:Poisson}

A departure point for the discussed
generalization of  linear QM
 is the observation that quantum theory can be
regarded as a particular {\it classical\/} infinite dimensional
Hamiltonian,
Poissonian or Nambu-like theory.

Let $\cal H$ be a Hilbert space. Consider the Hamilton equations
\begin{equation}
\omega^{AA'}(\alpha,\alpha')\frac{d\psi_A(\alpha)}{d\tau}=
\frac{\delta H}{\delta \psi^*_{A'}(\alpha')}
\end{equation}
and c.c.,
where the bars denote complex conjugations and the conventions
concerning primed and unprimed indices are assumed like in the
spinor abstract index calculus \cite{PR}. The summation
convention is as follows: We sum over repeated
Roman indices and integrate over repeated Greek ones.
The integration is with respect to some invariant, or
quasi-invariant measure on a finite dimensional manifold (mass
hyperboloid, spacelike hyperplane in the Minkowski space, etc.).
The symbol of the ``proper time"
 derivative describes a differentiation
with respect to a suitable foliation of space-time (Minkowskian
spacelike, or Galilean $t=$const hyperplanes, etc., see
Appendix).
In
Hilbertian formulation of QM the
``symplectic form" is given by the delta distribution
\begin{equation}
\omega^{AA'}(\alpha,\alpha'):=i  \delta^{AA'}\delta(\alpha ,
\alpha') =:\omega^{AA'}\delta(\alpha ,
\alpha')
\end{equation}
where $\delta^{AB'}=\delta_{AB'}=1$ if $A=B'$ and 0 for $A\neq
B'$ in the nonrelativistic QM. For the Dirac equation
$\delta^{AB'}$ and $\delta_{AB'}$ can be represented by the
Dirac matrix $\gamma_0$, and the Dirac delta function must
correspond to the choice of the spacelike hyperplane.
In the projective space formulation the symplectic form
corresponds to the Fubini-Study metric.
 The
inverse of $\omega^{AA'}(\alpha,\alpha')$ is
\begin{equation}
I_{AA'}(\alpha,\alpha'):=-i \delta_{AA'}\delta(\alpha,\alpha')=:
I_{AA'}\delta(\alpha ,\alpha')
\end{equation}
where by the inverse we understand that
\begin{eqnarray}
\omega^{AA'}(\alpha,\alpha') I_{BA'}(\beta,\alpha')&=&
\delta^A_B\delta(\alpha,\beta)\\
\omega^{AA'}(\alpha,\alpha') I_{AB'}(\alpha,\beta')&=&
\delta^{A'}_{B'}\delta(\alpha' , \beta').
\end{eqnarray}
Accordingly, the form of the Hamilton equations we shall use is
\begin{equation}
\frac{d\psi_A(\alpha)}{d\tau}=
I_{AA'}\frac{\delta H}{\delta \psi^*_{A'}(\alpha)}\label{Ham1}
\end{equation}
and c.c.
(\ref{Ham1})  describes a quantum evolution of
pure states. All observables of the linar theory
depend on $| \psi\rangle $ and $\langle \psi|$ {\em via\/}
 the density matrix
$\rho=| \psi\rangle \langle \psi|$.
 Let $F$ and $G$ be two such
observables, that is $F[\psi, \psi^*]=F[\rho]$ and
$G[\psi,
\psi^*]=G[\rho]$. The Poisson bracket resulting from the Hamilton
equations is
\begin{equation}
\{F,G\}=I_{AA'}\Bigl(
\frac{\delta F}{\delta\psi_{A}(\alpha)}
\frac{\delta G}{\delta \psi^*_{A'}(\alpha)}-
\frac{\delta G}{\delta\psi_{A}(\alpha)}
\frac{\delta F}{\delta \psi^*_{A'}(\alpha)}\Bigr).\label{Pois}
\end{equation}
Applying the chain rule to the components of the pure state
density matrix
\begin{equation}
\rho_{AA'}(\alpha,\alpha')=\psi_A(\alpha)
\psi^*_{A'}(\alpha')\label{pureDM}
\end{equation}
we find that
\begin{equation}
\{F,G\}=I_{AA'}\Bigl(
\frac{\delta
F}{\delta\rho_{AB'}(\alpha,\beta')}\rho_{CB'}(\gamma,\beta')
\frac{\delta
G}{\delta\rho_{CA'}(\gamma,\alpha)}- {\bigl(F\leftrightarrow
G\bigr)}
\Bigr).\label{Jbr}
\end{equation}
So long as the density matrix in (\ref{Jbr}) is given by
(\ref{pureDM}) the bracket is equivalent to the Poisson bracket
(\ref{Pois}). Jordan, in a context of the Weinberg's theory
\cite{Jordan} and for a finite dimensional Hilbert space,
investigated properties of the bracket (\ref{Jbr}) with $\rho$
being an arbitrary density matrix. For reasons that will be
explained below I will term such a general bracket the Bia\l
ynicki-Birula--Morrison--Jordan (BBMJ) bracket.

We will now show that (\ref{Jbr}), for a general $\rho$, can be
written in a form of a generalized Nambu bracket.  Let $\rho$ be
arbitrary. The BBMJ bracket can be rewritten as
\begin{equation}
\{F,G\}=
\rho_{AA'}(\alpha,\alpha')
\Omega^{AA'}_{{\ }{\ }{\ }BB'CC'}(\alpha,\alpha',
\beta,\beta',\gamma,\gamma')
\frac{\delta F}{\delta\rho_{BB'}(\beta,\beta')}
\frac{\delta G}{\delta\rho_{CC'}(\gamma,\gamma')}
\end{equation}
with
\begin{eqnarray}
\Omega^{AA'}_{{\ }{\ }{\ }BB'CC'}(\alpha,\alpha',
\beta,\beta',\gamma,\gamma')&=&
\delta^A_C\delta^{A'}_{B'}I_{BC'}\delta(\alpha,\gamma)
\delta(\alpha',\beta')\delta(\beta,\gamma')\nonumber \\
& &-
\delta^A_B\delta^{A'}_{C'}I_{CB'}\delta(\alpha,\beta)
\delta(\alpha',\gamma')\delta(\gamma,\beta')\nonumber\\
&=&
\Omega^{a}_{{\ }bc}
\end{eqnarray}
where, in analogy to the spinor calculus, we have clumped
together the respective quadruples of indices into composite ones
($a=(A,A',\alpha,\alpha')$, etc.).

The ``structure kernels" $\Omega^{a}_{{\
}bc}$
 satisfy conditions
characteristic for Lie-algebraic structure constants:
\begin{equation}
\Omega^{a}_{{\ }cb}=
-\Omega^{a}_{{\ }bc}
\label{antisymmetry}
\end{equation}
and
\begin{equation}
\Omega^{a}_{{\ }bc}
\Omega^{c}_{{\ }de}
+\Omega^{a}_{{\ }ec}
\Omega^{c}_{{\ }bd}
+\Omega^{a}_{{\ }dc}
\Omega^{c}_{{\ }eb}=0
\label{structure2}
\end{equation}
These two conditions imply the Jacobi identity. The composite
index form of the BBMJ bracket
\begin{equation}
\{F,G\}=
\rho_{a}
\Omega^{a}_{{\ }bc}
\frac{\delta F}{\delta\rho_{b}}
\frac{\delta G}{\delta\rho_{c}}\label{Mor br}
\end{equation}
shows that it takes the same form as the generalized BBM-Nambu
bracket written in terms of the Wigner function for a scalar
field \cite{Mor}. As a matter of fact, the BBMJ bracket is
simply a different representation of the BBM bracket.  The
formula (\ref{Mor br}) looks much the same as the Poisson
bracket related to the Kiryllow form on coadjoint
representations of Lie groups \cite{Arnold} (such
brackets for general structure constants are called the
Lie-Poisson brackets (cf. \cite{Karpacz})).

It remains to find out how to formulate the explicit {\it triple
bracket\/} equivalent to (\ref{Mor br}).

In order to do this we first have to define a ``metric tensor"
to lower the upper index in the structure kernels (Bia\l
ynicki-Birula and Morrison avoided this difficulty because the
field they considered had no spinor components). The
apparently natural guess (the Killing-Cartan metric)
\begin{equation}
g_{ab}
=
\Omega^{c}_{{\ }ad}
\Omega^{d}_{{\
}bc}
\label{zly}
\end{equation}
is incorrect as (\ref{zly}) involves expressions like
$\delta(0)$ which are not distributions in the Schwartz sense.

The correct definitions are
\begin{eqnarray}
g_{ab}
&=&
-I_{AB'}(\alpha,\beta')I_{BA'}(\beta,\alpha')\\
g^{ab}
&=&
-\omega^{AB'}(\alpha,\beta')\omega^{BA'}
(\beta,\alpha').
\label{MT}
\end{eqnarray}
The metric tensor is symmetric
\begin{equation}
g_{ab}=
g_{ba}
\end{equation}
and satisfies the invertibility conditions
\begin{equation}
g^{ab}
g_{bc}
=
g_{cb}
g^{ba}
=\delta^A_C\delta^{A'}_{C'}\delta(\alpha,
\gamma)\delta(\alpha',\gamma')
=:
\delta^a_c.
\end{equation}
The metric tensor is a useful tool. Consider for example a
$\rho$-independent $F_b=F_{BB'}(\beta,\beta')$.
Then
\begin{equation}
F[\rho]=g^{ab}\rho_a
F_b
=
\rho^{B'}_{{\phantom{A}}A'}(\beta',\alpha')
F^{A'}_{{\phantom{A}}B'}(\alpha',\beta')={\rm Tr}\,\rho\hat F
\end{equation}
and we see that linear observables can be naturally expressed
with the help of (\ref{MT}). This example is important also as
an illustration of the convention concerning lowering and
raising of indices. For notice that
\begin{equation}
\frac{\delta F}{\delta\rho^{B'}_{{\phantom{A}}A'}
(\beta',\alpha')}=
F^{A'}_{{\phantom{A}}B'}(\alpha',\beta')\label{operator}
\end{equation}
although the staggering of indices like
$F_{B'}^{{\phantom{A}}A'}(\beta',\alpha')$ might seem more
natural.

The fully covariant form of the structure kernels is
\begin{equation}
\Omega_{abc}
= - I_{AB'}(\alpha,\beta')I_{CA'}(\gamma,\alpha')I_{BC'}
(\beta,\gamma')
 + I_{AC'}(\alpha,\gamma')I_{BA'}(\beta,\alpha')I_{CB'}
(\gamma,\beta').
\end{equation}
One easily verifies that
$\Omega_{abc}
$ is totally
antisymmetric.

Following Bia\l ynicki-Birula and Morrison let us introduce the
functional
\begin{equation}
S_2={1\over 2} g^{ab}
\rho_a
\rho_b=
{1\over 2}{\rm Tr}\,(\rho^2),
\end{equation}
which is one half of the inverse of
R\'enyi's $2^*$-entropy.

The BBMJ bracket is now equal to the following triple bracket
\begin{equation}
\{F,G\}=[F,G,S_2]=
\Omega_{abc}
\frac{\delta F}{\delta\rho_{a}}
\frac{\delta G}{\delta\rho_{b}}
\frac{\delta S_2}{\delta\rho_{c}}.
\end{equation}
The antisymmetry of the triple bracket means that $S_2$ is the
Casimir for the BBMJ bracket Lie algebra of observables. Another
Casimir is
${\rm Tr}\,\rho$
because $\{{\rm Tr}\,\rho,F\}=0$ for {\em any\/} differentiable
$F$ (hence not only linear).
 The wave functions have been eliminated from the
dynamical equations, but the Hilbert space background is
implicitly present in the structure kernels and the metric
tensor which are defined in terms of $\omega$ and $I$, and in
the very notion of the density matrix which acts in the Hilbert
space.

Components of the pure state density matrix satisfy
\begin{equation}
\frac{d}{d\tau}\rho_{a}=
\{\rho_{a},
H\}.\label{Liouville}
\end{equation}
which holds also for general density matrices as can be seen
from the familiar, operator version of the Liouville--von
Neumann equation. It follows that the density matrices form a
Poisson manifold, as opposed to state vectors that form a phase
space.

\section{Nonlinear Quantum Mechanics as a Generalized Nambu Mechanics}

The generalizations of quantum mechanics considered by Kibble
\cite{Kibble} and Weinberg \cite{W2} are based on the
Hamiltonian framework. The nonlinear evolution is introduced
through an extension of the class of admissible Hamiltonian
functions. More generally, all canonical transformations are
generated by a larger class of functionals on Hilbert or
projective spaces. The functionals are a generalization of averages
of observable quantities. This fact leads to the fundamental
difficulty in constructing a probability interpretation of such
theories:
The generalized observables do not form an associative
algebra which makes impossible a unique definition of powers of
observables,
the formal counterpart of higher moments of random variables
measured in experiments.

The triple bracket form of the Liouville-von Neumann equation
shows that the time evolution in linear QM has, in fact, {\em two \/}
generators: the average energy (Hamiltonian function) and the
Casimir $S$,
which measures R\'enyi's $\alpha=2$ entropy (or, even more
directly, Dar\'oczy entropy of order 2). It is natural to
ask what will be
changed in the theory if, instead of generalizing the class of
admissible Hamiltonian functions, we shall extend the class of
entropies. A physical meaning of such an extension would be the
one required by Wigner in his paradox of a friend: We extend
quantum mechanics to systems that can gain information.
The extended theory has a well defined probability
interpretation, because the observables are represented by linear
operators, provided the scaling by a constant,
$\rho\to\lambda\rho$, is a symmetry of the dynamics. This
imposes on the generalized entropies the 2-homogeneity
condition: $S(\lambda\rho)=\lambda^2 S(\rho)$.

Only for $S[\rho]=1/2 {\rm Tr}(\rho^2)$
the linear observables are closed under the action
of the bracket $\{\cdot,\cdot\}_S:=[\cdot,\cdot,S]$. If we
extend the class of acceptable
$S$, we have to accept also a somewhatOB stronger form of the
complementarity principle than in linear QM: Observables are
always complementary
to their (nonvanishing) time derivatives
(see Sec.~\ref{Comments}). We shall
begin the discussion of the generalization with the question
whether, for general $S$, the manifold of states is the Poisson
manifold.

\subsection{The Jacobi Identity}

Let $F$, $G$, $H$ and $S$ be arbitrary twice functionally
differentiable functionals. We consider the expression
\begin{eqnarray}
J&=&\bigl\{\{F,G\}_S,H\bigr\}_S+\bigl\{\{H,F\}_S,G\bigr\}_S
+\bigl\{\{G,H\}_S,F\bigr\}_S
\nonumber\\
&=&
\frac{\delta F}{\delta\rho_{d}}
\frac{\delta G}{\delta\rho_{e}}
\frac{\delta^2 S}{\delta\rho_{a}\delta\rho_{f}}
\frac{\delta H}{\delta\rho_{b}}
\frac{\delta S}{\delta\rho_{c}}\bigl(
\Omega_{def}\Omega_{abc}+\Omega_{bdf}\Omega_{aec}
+\Omega_{ebf}\Omega_{adc}
\bigr)\label{structure2'}
\end{eqnarray}
which holds good for any $S$.  $\frac{\delta^2
S}{\delta\rho_{a}\delta\rho_{f}}=g^{af}$ for $S=S_2$ and
(\ref{structure2'}) vanishes in virtue of (\ref{structure2}).
For more general $S=S(f_2[\rho])$ we find
\begin{eqnarray}
\frac{\delta S}{\delta\rho_{c}}&=& 2\frac{\partial S}{\partial
f_2}\rho^c\\
\frac{\delta^2 S}{\delta\rho_{a}\delta\rho_{f}}&=&
4 \frac{\partial^2 S}{\partial f_2^2}\rho^a\rho^f+
2\frac{\partial S}{\partial f_2}g^{af}.
\end{eqnarray}
Inserting these expressions into (\ref{structure2'}) we obtain
\begin{equation}
J=8\frac{\delta F}{\delta\rho_{d}}
\frac{\delta G}{\delta\rho_{e}}
\frac{\partial^2 S}{\partial f_2^2}\rho^a\rho^f
\frac{\delta H}{\delta\rho_{b}}
\frac{\partial S}{\partial f_2}\rho^c\bigl(
\Omega_{def}\Omega_{abc}+\Omega_{bdf}\Omega_{aec}
+\Omega_{ebf}\Omega_{adc}
\bigr)=0
\end{equation}
since $\Omega_{abc}\rho^a\rho^c=0$. With this choice of $S$ we
obtain the dynamics given by
\begin{equation}
\frac{d}{d\tau}\rho_{a}=
\{\rho_{a},
H\}_{S_2} C[\rho]
\end{equation}
where $C[\rho]=2\frac{\partial S}{\partial f_2}=C(f_2[\rho])$ is
an integral of motion, as we shall see later. The only
difference with respect to ordinary QM would be in a
$\rho$-dependent rescaling of time, a phenomenon that, in
principle, might influence lifetime characteristics of physical
processes.

For more general $S$ the question of the Jacobi identity is
open, hence we have to accept the possibility that
mixed states in the generalized QM do not form a Poisson
manifold. This would not be surprising, since in various versions
of generalizations of the Nambu mechanics, the Jacobi identity
does not hold.

\subsection{Composite Systems in the New Framework}

Let the Hilbert space in question and the density matrix of some
composite system be ${\cal H}={\cal H}_1\otimes{\cal H}_2$ and
\begin{equation}
\rho_a=\rho_{AA'}(\alpha,\alpha')=
\rho_{A_1A_2A'_1A'_2}
(\alpha_1,\alpha_2,\alpha'_1,\alpha'_2).
\end{equation}
The same doubling of indices concerns
\begin{equation}
I_{AA'}(\alpha,\alpha')=-i \delta_{A_1A'_1}\delta_{A_2A'_2}
\delta(\alpha_1,\alpha'_1)\delta(\alpha_2,\alpha'_2).
\end{equation}
Reduced density matrices of the two subsystems are
\begin{eqnarray}
\rho^I_{A_1A'_1}(\alpha_1,\alpha'_1)&=&
\delta^{A_2A'_2}\delta(\alpha_2,\alpha'_2)\rho_{A_1A_2A'_1A'_2}
(\alpha_1,\alpha_2,\alpha'_1,\alpha'_2) \label{rdmI}\\
\rho^{II}_{A_2A'_2}(\alpha_2,\alpha'_2)&=&
\delta^{A_1A'_1}\delta(\alpha_1,\alpha'_1)\rho_{A_1A_2A'_1A'_2}
(\alpha_1,\alpha_2,\alpha'_1,\alpha'_2)\label{rdmII}
\end{eqnarray}
and satisfy
\begin{equation}
\frac{\delta\rho^I_{A_1A'_1}(\alpha_1,\alpha'_1)}{\delta
\rho_{B_1B_2B'_1B'_2}
(\beta_1,\beta_2,\beta'_1,\beta'_2)}=
\delta^{B_1}_{A_1}\delta^{B_2B'_2}\delta^{B'_1}_{A'_1}
\delta(\beta_1,\alpha_1)\delta(\beta_2,\beta'_2)
\delta(\beta'_1,\alpha'_1)\label{RDM1}
\end{equation}
and
\begin{equation}
\frac{\delta\rho^{II}_{A_2A'_2}(\alpha_2,\alpha'_2)}{\delta
\rho_{B_1B_2B'_1B'_2}
(\beta_1,\beta_2,\beta'_1,\beta'_2)}=
\delta^{B_2}_{A_2}\delta^{B_1B'_1}\delta^{B'_2}_{A'_2}
\delta(\beta_2,\alpha_2)\delta(\beta_1,\beta'_1)
\delta(\beta'_2,\alpha'_2).\label{RDM2}
\end{equation}
The structure kernels for the composite system are
\begin{eqnarray}
\Omega_{abc}&=&
\Omega_{a_1a_2b_1b_2c_1c_2}\nonumber\\
&=&-i \Bigl(
\delta_{A_1B'_1}\delta_{C_1A'_1}\delta_{B_1C'_1}
\delta_{A_2B'_2}\delta_{C_2A'_2}\delta_{B_2C'_2}
\times\nonumber\\
&\phantom{=}&\phantom{-i
\Bigl(}\delta(\alpha_1,\beta'_1)\delta(\gamma_1,\alpha'_1)\delta
(\beta_1,\gamma'_1)
\delta(\alpha_2,\beta'_2)\delta(\gamma_2,\alpha'_2)\delta
(\beta_2,\gamma'_2)\nonumber\\ &\phantom{=}&\phantom{i \Bigl(}-
\delta_{A_1C'_1}\delta_{B_1A'_1}\delta_{C_1B'_1}
\delta_{A_2C'_2}\delta_{B_2A'_2}\delta_{C_2B'_2}\times
\nonumber\\
&\phantom{=}{}&\phantom{-i \Bigl(}
\delta(\alpha_1,\gamma'_1)\delta(\beta_1,\alpha'_1)\delta
(\gamma_1,\beta'_1)
\delta(\alpha_2,\gamma'_2)\delta(\beta_2,\alpha'_2)\delta
(\gamma_2,\beta'_2)\Bigr).\nonumber\\
\label{strucomp}
\end{eqnarray}
The following two results solve generally the question of
faster-than-light telegraphs in both Hamiltonian and triple
bracket frameworks.

\begin{lem}
\label{lemma}
Reduced density matrices of the subsystems satisfy
\begin{equation}
\Omega_{abc}
\frac{\delta\rho^I_{d_1}}{\delta
\rho_{a}}
\frac{\delta\rho^{II}_{d_2}}{\delta
\rho_{b}}=0.
\end{equation}
\end{lem}
{\it Proof\/}: It is sufficient to contract (\ref{strucomp})
with (\ref{RDM1}) and (\ref{RDM2}).$\Box$

\begin{th}
\label{Th 7.1}
Let $F=F[\rho^I]$ and $G=G[\rho^{II}]$, that is depend on $\rho$
{\it via\/} {\em (\ref{rdmI})} and {\em (\ref{rdmII})}, then for
any $S$
\begin{equation}
\{F,G\}_S=0.
\end{equation}
\end{th}
{\it Proof\/}: By virtue of the lemma one has
\begin{equation}
0=
\Omega_{abc}
\frac{\delta\rho^I_{d_1}}{\delta
\rho_{a}}
\frac{\delta\rho^{II}_{d_2}}{\delta
\rho_{b}}
\frac{\delta F}{\delta\rho^I_{d_1}}
\frac{\delta G}{\delta\rho^{II}_{d_2}}
\frac{\delta S}{\delta\rho_{c}}=\{F,G\}_S.
\end{equation}
$\Box$

Notice that we have not assumed anything but differentiability
not only about $S$ but also about $F$ and $G$. So, in
particular, for arbitrary (nonlinear) observables and $S=S_2$ we
obtain the Polchinski-Jordan result for Weinberg's nonlinear QM.

\subsection{Density Matrix Interpretation of Solutions
of the Generalized Evolution Equation}

One of the essential questions we have to clarify concerns the
density matrix interpretation of the solutions of
the generalized Liouville-von Neumann equation
\begin{equation}
\frac{d}{d\tau}\rho_a=[\rho_a,H,S].\label{S-Liouville}
\end{equation}
There is no general {\it
a priori\/} guarantee that the generalized dynamics will
conserve positivity of $\rho$. The next theorems will give a
partial answer to this problem.

In order to attack the question we have to make the language of
the $S$-brackets more readable. Consider the triple bracket
$[F,G,H]$ of arbitrary functionals $F$, $G$ and $H$. We find
that
\begin{eqnarray}
i [F,G,H]&=&
\frac{\delta F}{\delta\rho^{B'}_{{\phantom{A}}A'}(\beta',\alpha')}
\frac{\delta G}{\delta\rho^{C'}_{{\phantom{A}}B'}(\gamma',\beta')}
\frac{\delta
H}{\delta\rho^{A'}_{{\phantom{A}}C'}(\alpha',\gamma')}
\nonumber\\
&\phantom{.}&\phantom{\delta\rho^{B'}_{{\phantom{A}}A'}}-
\frac{\delta
F}{\delta\rho^{C'}_{{\phantom{A}}A'}(\gamma',\alpha')}
\frac{\delta
G}{\delta\rho^{A'}_{{\phantom{A}}B'}(\alpha',\beta')}
\frac{\delta
H}{\delta\rho^{B'}_{{\phantom{A}}C'}(\beta',\gamma')}.
\label{FGH}
\end{eqnarray}
Applying the notation of (\ref{operator}) (where now the
``operator" kernels are in general $\rho$-dependent) we
transform (\ref{FGH}) into
\begin{eqnarray}
F_{{\phantom{A}}B'}^{A'}(\alpha',\beta') {
G}_{{\phantom{A}}C'}^{B'}(\beta',\gamma') {
H}_{{\phantom{A}}A'}^{C'}(\gamma',\alpha')
\phantom{{ F}_{{\phantom{A}}C'}^{A'}(\alpha',\gamma')
{ G}_{{\phantom{A}}A'}^{B'}(\beta',\alpha') {
H}_{{\phantom{A}}B'}^{C'}(\gamma',\beta')}
\nonumber\\
-\, { F}_{{\phantom{A}}C'}^{A'}(\alpha',\gamma') {
G}_{{\phantom{A}}A'}^{B'}(\beta',\alpha') {
H}_{{\phantom{A}}B'}^{C'}(\gamma',\beta') ={\rm Tr}\, \bigl(
[\hat F, \hat G] \hat H \bigr).
\label{NLoper}
\end{eqnarray}
In the last line we have introduced an abbreviated convention
based on the assignment to any functional $F$ of an operator
\begin{equation}
\hat F=\frac{\delta F}{\delta \rho}
\end{equation}
which is defined by the kernel form used in (\ref{NLoper}). For
example
\begin{equation}
\rho=\frac{\delta S_2}{\delta \rho},
\end{equation}
and
\begin{equation}
\frac{\delta {\rm Tr}\, \bigl(\rho^n\bigr)}{\delta
\rho}=n\rho^{n-1},
\end{equation}
the latter being the shortened form of
\begin{eqnarray}
\frac{\delta {\rm Tr}\, \bigl(\rho^n\bigr)}{\delta
\rho_{AA'}(\alpha,\alpha')}&=&
n\delta^{B_n'A}\delta^{A'B_2}\delta^{B_2'B_3}
\dots\delta^{B_{n-1}'B_n}
\nonumber\\
&\phantom{=}&\quad
\times\,\delta(\beta_n',\alpha)
\delta(\alpha',\beta_2)\delta(\beta_2',\beta_3)
\dots\delta(\beta_{n-1}',\beta_n)\nonumber\\
&\phantom{=}&\quad
\times\,\rho_{B_2B_2'}(\beta_2,\beta_2')\dots\rho_{B_nB_n'}
(\beta_n,\beta_n').
\end{eqnarray}
The first of these implies the known result
\begin{equation}
[F,G,S_2]=-i {\rm Tr}\,\bigl( \rho [\hat F,\hat G]\bigr)
\end{equation}
leading to the von Neumann/Heisenberg equations for
states/observables in linear QM
\begin{equation}
\frac{d}{d\tau}{\rm Tr}\bigl(\rho\hat F\bigr)=
-i {\rm Tr}\,\bigl( \rho [\hat F,\hat H]\bigr).
\end{equation}
The same equation is valid in the Polchinski-Jordan density
matrix formulation of Weinberg's NLQM \cite{Polchinski,Jordan},
but then $\hat F=\hat
F[\rho]$, etc.
Consider now a functional $S$ (differentiable in $f_k$)
\begin{equation}
S[\rho]=S\bigl(f_1[\rho],\dots,f_n[\rho],\dots\bigr)\label{S'}
\end{equation}
where $f_k[\rho]= {\rm Tr}\, (\rho^k)$.

\begin{th}
For any $m\in {\bf N}$, and any $G$, if $S$ satisfies {\em
(\ref{S'})} then
\begin{equation}
[f_m, G, S]=0.
\end{equation}
\end{th}
{\it Proof\/}:
\begin{eqnarray}
[{\rm Tr}\, (\rho^m), G, S]=\sum_n[{\rm Tr}\, (\rho^m), G,
f_n]\frac{\partial S}{\partial f_n}=-i m\sum_n n{\rm Tr}\,\bigl(\hat G
[\rho^{m-1},\rho^{n-1}]\bigr)\frac{\partial S}{\partial f_n}=0.
\end{eqnarray}
$\Box$
This interesting result covers many  nontrivial
generalizations of $S_2$. As a by-product it shows also that the
same property holds for the Weinberg-Polchinski-Jordan NLQM
because we have not assumed that $G$ is
linear in $\rho$ (moreover, it includes other theories where
observables do not satisfy any homogeneity condition). The
particular case $m=1$ implies that ${\rm Tr}\,\rho$ is conserved
by all evolutions, a fact important for a definition of
averages. For pure states ${\rm Tr}\, (\rho^m)=({\rm
Tr}\,\rho)^m$ so that the integrals $f_m$ are not necessarily
independent, but for all $m,n$ $f_m$ and $f_n$ are in involution
with respect to $\{\cdot,\cdot\}_S$. Jordan proved in
\cite{Jordan} by an explicit calculation that in his
formulation of Weinberg's nonlinear QM ${\rm Tr}\,\rho$ and
${\rm Tr}\,\rho^2$ are conserved --- our theorem considerably
generalizes this result.

\begin{th}
Let $S$ satisfy {\em (\ref{S'})} and $\rho_t$ be a self-adjoint
 solution of
{\em (\ref{S-Liouville})}. If $\rho_0$ is positive and has a
finite number of nonvanishing eigenvalues $p_k(0)$,
$0<p_k(0)\leq 1$, then the eigenvalues of $\rho_t$ are integrals
of motion, and the evolution conserves positivity of $\rho_t$.
\end{th}
{\it Proof\/}: Since the nonvanishing eigenvalues of $\rho_0$
satisfy $0<p_k(0)\leq 1<2$, it follows that for any $\alpha$
$p_k(0)^\alpha$ can be written in a form of a convergent Taylor
series. By virtue of the spectral theorem the same holds for
$\rho_0^\alpha$ and ${\rm Tr}\, (\rho_0^\alpha)$. Each element
of the Taylor expansion of ${\rm Tr}\, (\rho_0^\alpha)$ is
proportional to $f_n[\rho_0]$, for some $n$. But
$f_n[\rho_0]=f_n[\rho_t]$ hence
\begin{equation}
{\rm Tr}\, (\rho_0^\alpha)={\rm Tr}\, (\rho_t^\alpha)=\sum_k
p_k(0)^\alpha=
\sum_k p_k(t)^\alpha
\end{equation}
for all real $\alpha$. Since all $p_k(0)$ are assumed to be
known (the initial
condition), we know also $\sum_k p_k(0)^\alpha=\sum_k
p_k(t)^\alpha$ for any $\alpha$. We can now apply the  result
used in the information theory \cite{Renyi} stating that the
knowledge of $\sum_k p_k(t)^\alpha$ for all $\alpha$ {\it
uniquely\/} determines $p_k(t)$. The continuity in $t$ implies
that $p_k(t)=p_k(0)$. $\Box$

The spectral decomposition of the density matrix
\begin{equation}
\rho_t=\sum_k p_k | {k,t}\rangle\langle {k,t}|,
\end{equation}
where $t\mapsto| {k,t}\rangle$ defines a one-parameter continuous
family of orthonormal vectors, leads to the unitary (although
$\rho$-dependent) transformation $|
{k,t}\rangle=U(\rho_t,\rho_0)| {k,0}\rangle$. The density matrix evolves
then as follows
\begin{equation}
\rho_t= U(\rho_t,\rho_0)\rho_0 U(\rho_t,\rho_0)^{-1}.
\end{equation}
The question whether the same holds good for $\rho_0$ having an
infinite number of nonvanishing eigenvalues will be left open
here. In any case, it seems that the above theorem is sufficient
at least ``for all practical purposes".

To make our proposal  more concrete, we have to
choose some explicit ``physical" class of $S$ --- and here the
information theoretic introduction may be helpful.

The suggestion of Wigner that a natural arena for nonlinear
generalizations of the linear formalism of QM is the domain of
{\it observations\/} leads to investigation of systems that {\it
can gain information\/} hence are described by $\alpha\neq 2$
entropies. A homogeneity preserving generalization of $S_2$ for
other $\alpha$-entropies can be, for instance,
\begin{equation}
S_\alpha[\rho]=\Bigl(1-\frac{1}{\alpha}\Bigr)
\frac{\bigl({\rm Tr}\,(\rho^\alpha)\bigr)^{1/(\alpha-1)}}{({\rm
Tr}\,\rho)
^{1/(\alpha-1)-1}}.\label{S-alpha}
\end{equation}
The choice of the denominator is important only from the point
of view of the homogeneity of the evolution equation. The
multiplier $1-1/\alpha$ guarantees that the evolution of pure
states is the same, hence {\it linear\/}, for all $\alpha$ (
this is reasonable as pure states have the same, vanishing
$\alpha$-entropies). The generalized Liouville-von Neumann
equation following from (\ref{S-alpha}) is
\begin{equation}
i\frac{d}{d\tau}\rho=\frac{\bigl({\rm
Tr}\,(\rho^\alpha)\bigr)^{1/(\alpha-1)-1}} {({\rm Tr}\,\rho)
^{1/(\alpha-1)-1}}[\hat H,\rho^{\alpha-1}].
\end{equation}
For pure states and ${\rm Tr}\,\rho=1$, $\rho^n=\rho$ and the
equation reduces to the ordinary, linear one; for mixed states
the evolution is nonlinear unless the states are ``so mixed"
that $\rho$ is proportional to the unit operator (which makes
sense in finite dimensional cases, of course) and all
$\alpha$-entropies reduce to the Hartley formula.

The evolution of (now linear) observables is governed by
\begin{equation}
i\frac{d}{d\tau}F=\frac{\bigl({\rm
Tr}\,(\rho^\alpha)\bigr)^{1/(\alpha-1)-1}} {({\rm Tr}\,\rho)
^{1/(\alpha-1)-1}}{\rm Tr}\,\bigl(\rho^{\alpha-1}[\hat F,\hat
H]\bigr)
\end{equation}
which shows that for the generalized $S$ the time derivative of
an observable is not linear in the density matrix.  For
$\alpha=2$ the equations reduce again to the ordinary linear
equations.

It seems that the following choice of $S_\alpha$ is also
interesting:
\begin{equation}
S_\alpha[\rho]={1\over 2}
\frac{\bigl({\rm Tr}\,(\rho^\alpha)
\bigr)^{1/(\alpha-1)}}{({\rm Tr}\,\rho)
^{1/(\alpha-1)-1}}.
\end{equation}
For pure states the expression reduces to the linear form
${1\over 2}
\langle\psi|\psi\rangle^2={1\over 2}{\rm Tr}\,(\rho^2)$.
The density matrix would satisfy then the equation
\begin{equation}
i\frac{d}{d\tau}\rho= {1\over 2}\frac{\alpha}{\alpha-1}
\frac{\bigl({\rm Tr}\,(\rho^\alpha)\bigr)^{1/(\alpha-1)-1}}
{({\rm Tr}\,\rho) ^{1/(\alpha-1)-1}}[\hat
H,\rho^{\alpha-1}]\label{S-alpha'}
\end{equation}
which for pure states and normalized $\rho$ would become
\begin{equation}
2\frac{\alpha -1}{\alpha}i\frac{d}{d\tau}\rho= [\hat H,\rho]
\end{equation}
and the ``Boltzmann-Shannon classical limit" $\alpha\to 1$ of
the R\'enyi entropy is indistinguishable from the $\hbar\to 0$
classical
limit of QM.

\subsection{
Composition Problem for Subsystems with Different Entropies}

Assuming that the formalism is applicable to a description of
the composite ``object+observer" system, where the nonlinearity
is a feature of the
{\em observer\/}, we have to know how
to combine systems that are described by different entropies.

I think it is best to approach the question again in an
information theoretic way. To begin with, let us consider a
system whose entropy is $I_{\alpha}$, and whose subsystems have
entropies of the {\it same\/} kind: for all $k$ a $k$-th
system's entropy satisfies $I_{\alpha_k}=I_{\alpha}$.  Let the
$k$-th subsystem be described by a reduced density matrix
$\rho_k$. The entropy of the ``large" system should be defined,
as usual in information theory, as the average entropy of the
subsystems.  The overall entropy of the large systems should not
depend on the way we decompose it into subsystems. Therefore the
average cannot have the apparently natural form
\begin{equation}
I_{{\alpha_1}\dots {\alpha_n}}[\rho]=\sum_{k}^n p_k
I_{\alpha}[\rho_k],\label{7.100}
\end{equation}
where $p_k$ are some weights, because the LHS is sensitive to
correlations between the subsystems whereas the RHS is not, so
that the entropy would be sensitive to the decompositions which
are arbitrary.  It seems we have to assume that in such a case
the composition takes the trivial form
\begin{equation}
I_{{\alpha_1}\dots {\alpha_n}}[\rho]=\sum_{k}p_k I_{\alpha_k}[\rho]=
\sum_{k}p_k
I_{\alpha}[\rho]= I_{\alpha}[\rho].\label{comp entropy}
\end{equation}
Consider now a situation where the different subsystems have
{\it different\/} entropies, say, $I_{\alpha_k}$. The average
entropy of the composite system is now defined in analogy to
(\ref{comp entropy}) as
\begin{equation}
I_{\alpha_1\dots\alpha_n}[\rho]=\sum_{\alpha_k}p_k
I_{\alpha_k}[\rho] \label{comp entropy'}
\end{equation}
where the probabilities $p_k$ are weights describing the
``percentage" of each of the entropies in the overall entropy of
the system. We do not know how to determine the weights --- they
can play a role of parameters characterizing the system.

The above definitions imply that the $\alpha^*$-entropy of the
large system is
\begin{equation}
I^*_{\alpha_1\dots\alpha_n}[\rho]=\prod_{\alpha_k}
I^*_{\alpha_k}[\rho]^{p_k},\label{7.103}
\end{equation}
so it is natural to define
\begin{equation}
S_{\alpha_1\dots\alpha_n}[\rho]=\prod_{\alpha_k}
S_{\alpha_k}[\rho]^{p_k}.\label{7.104}
\end{equation}
Denoting the latter expression by $S$, we obtain
\begin{equation}
\frac{d}{d\tau}\rho_b=
\sum_{\alpha_k}p_k [\rho_b,H,S_{\alpha_k}]
\frac{S}{S_{\alpha_k}}.
\end{equation}
Consider again a system which consists of subsystems equipped
with the entropy of the same kind. Then
$S_{\alpha_k}=S_{\alpha_l}=S$ for all $k$ and $l$ and the system
evolves according to
\begin{equation}
\frac{d}{d\tau}\rho_b=[\rho_b,H,S]
\end{equation}
as expected.

The next possibility is that the entropies that sum to the overall
entropy are again sums of some other entropies. The description
of the whole system should not depend on the order in which the
partial entropies are summed up. So consider two entropies
$S^{I}$ and $S^{II}$, with appropriate weights $\lambda^{I}$ and
$\lambda^{II}$, and let the entropies $S^I$ and $S^{II}$ consist
of some other entropies $S^{I}_k$ and $S^{II}_l$ appearing with
weights $\{p_k^I\}_{k=1}^N$ and $\{p_l^{II}\}_{l=1}^M$,
respectively. Then
\begin{equation}
\frac{d}{d\tau}\rho_b=
\sum_k\lambda^{I}p_k^I\{\rho_b,H\}_{S^I_k}
\frac{S}{S^I_k}
+
\sum_l\lambda^{{II}}p_l^{{II}}\{\rho_b,H\}
_{S^{II}_l}
\frac{S}{S^{II}_l}
\end{equation}
which shows that the evolution can be indeed consistently
composed of ``sub-entropies".

If all $S_{\alpha_k}[\rho]$ depend on $\rho$ only {\it via\/}
$f_m[\rho]$, like in our definitions (\ref{S-alpha}) and
(\ref{S-alpha'}), we know that on general grounds they are
integrals of motion. Consider a subsystem described by $\rho_k$
and which is noninteracting with the subsystem ``where the
nonlinearity resides". In such a case the overall Hamiltonian
function is
$
H[\rho]=\sum_k H_k[\rho_k]
$
and
\begin{equation}
\frac{d}{d\tau}\rho_{k\,b}=\sum_l
p_l [\rho_{k\,b},H_k[\rho_k],S_{\alpha_l}]
\frac{S}{S_{\alpha_l}}.
\end{equation}
A system described by $\rho_k$ and $S_{\alpha_k}$ can be totally
isolated from the ``rest of the Universe", if for $k\neq l$,
$
[\rho_{k\,b},H_k[\rho_k],S_{\alpha_l}]=0
$
and
$p_k=S_{\alpha_k}/S.$
However, even in such a case the global properties of the large
system leave their mark on the local properties of all the
subsystems as
$
p_k\,S/S_{\alpha_k}
$
is at most an integral of motion hence depends on initial
conditions.  Consider a general $H$ (including the interaction)
and let
\begin{equation}
S[\rho]=S_{\alpha_1\dots\alpha_n}[\rho]=\prod_{\alpha_k}
S_{\alpha_k}[\rho_k]^{p_k}.\label{7.108}
\end{equation}
where the different subsystems have different entropies.  Each
of the sub-entropies satisfies
\begin{equation}
\frac{d}{d\tau}S_{\alpha_k}[\rho_k]
=\sum_{\alpha_l}p_l
[S_{\alpha_k}[\rho_k],H,S_{\alpha_l}(\rho_l)]
\frac{S}{S_{\alpha_l}}=0
\end{equation}
in virtue of the theorem~\ref{Th 7.1} (we have used here the
fact that $\{F,H\}_S=-\{F,S\}_H$), and the antisymmetry of the
triple bracket. Therefore not only the overall entropy, but also
the sub-entropies are integrals of motion {\it even if the
Hamiltonian function $H$ contains interaction terms\/}.

Consider now the reduced density matrix $\rho_k$ of one of the
subsystems. Using the same theorem we find that
\begin{equation}
\frac{d}{d\tau}\rho_{k\,b}=
p_k [\rho_{k\,b},H,S_{\alpha_k}]
\frac{S}{S_{\alpha_k}}.\label{7.110}
\end{equation}
where $S/S_{\alpha_k}$ is an integral of motion but its value
depends on initial conditions. If the change of the initial
conditions does not affect the reduced density matrices in
(\ref{7.108}), the integral of motion is also unchanged.
Therefore in order to change this quantity we have to change
correlations between the subsystems. In particular, a time
dependence of a linear system which is noninteracting with the
nonlinear one is insensitive to changes of initial conditions
within the linear system if the particular form (\ref{7.108})
holds. For global entropies different from (\ref{7.108}) some
kind of sensitivity appears but the influences between the
subsystems cannot propagate faster than light unless we
introduce the projection postulate.

Such a trace of nonlinearity observed in some linear system
might be used to detect the nonlinearity. Following Santilli
\cite{Santilli} we can
expect that an evolution of an internal part of a hadron may be
nonliner (like in hadronic mechanics). In such a case
correlations between a hadron (say, a proton) and some linear
system (say, an electron) could be observed in a form of a
$\rho$-dependent rescaling of time in the electron's evolution.

\section{Comments}
\label{Comments}

S.~Weinberg wrote in \cite{Dreams} that the ``theoretical
failure to find a plausible alternative to quantum
mechanics, even more than the precise experimental verification of
linearity, suggests (...) that quantum mechanics is the way it is
because any small change in quantum mechanics would lead to logical
absurdities. If this is true, quantum mechanics may be a permanent part
of physics. Indeed, quantum mechanics may survive not merely as an
approximation to a deeper truth,
(...) but as a precisely valid feature of the final theory."
This kind of conviction followed from the
internal theoretical difficulties of the generalizations based
on nonlinear Schr\"odinger equations and general Hamiltonian
framework. These difficulties have been discussed in detail in
\cite{I}.

The proposal based on the generalized Nambu dynamics is free
from those difficulties. However, it has new features with
respect to ordinary QM. One, the stronger complementarity
principle, has already been  announced. In the generalized
framework a time derivative of an observable will not, in
general, be linear in the density matrix. Since we have defined
observables as functions necessarily linear in $\rho$, the time
derivatives of observables are not themselves observables.

I propose the following interpretation of this fact. To focus our
attention let us consider linear QM and the
nonrelativistic position operator. An average velocity of an
ensemble of particles can be calculated either by first calculating an
average position and then taking its time derivative, or by
first measuring the velocity of each single particle and then
taking the average. We can say that the first procedure is a
calculation of the time derivative of an average, whereas the
latter is taking the average of the time derivative.
The situation can be  described symbolically by the equation
\begin{equation}
\frac{d}{dt}\langle \vec q\rangle =\langle\frac{d}{dt}\vec
q\rangle .\label{q-v}
\end{equation}
An important property of  QM is the impossibility of
realizing the two procedures simultaneously, as $\vec
v=\vec p/m$ and $\vec q$ are {\it complementary\/}. It follows
that in a concrete experiment we have to decide which way of
measuring to choose. In this meaning if we can measure $\vec q$,
we cannot measure $\frac{d}{dt} \vec q$, and {\it vice versa\/}.
To express it differently, if $\vec q$ is observable (not {\it
an\/} observable!) then
$\frac{d}{dt} \vec q$ is not.

In triple bracket NLQM the observables will be defined as
quantities that are in
one-to-one relationship to some experimentally measured random
variables (hence the linearity in $\rho$). Two observables will
be said to be complementary if
there does not exist a physical situation where the two
respective random variables can be measured {\it
simultaneously\/}, that is, in a single run of an experiment.
There can exist linear operators representing the position and
the velocity of single members of an ensemble, but if the
ensemble evolves in a nonlinear way, the averages of those
observables do not have to satisfy the inherently linear
condition (\ref{q-v}), if the experimental procedures necessary
for their measurements cannot be simultaneously realized.

Another fundamental problem, arising in the Nambu-like
description, is the action principle leading to the
triple bracket equation. The Hamiltonian NLQM proposed by Kibble
or Weinberg can be derived from the ordinary Lagrangian
formalism. The triple bracket form of dynamics must follow from
a new kind of variational principle.
The variational principle  proposed
recently by Takhtajan \cite{Tak} suggests an interesting
direction for further investigations.

\section{Appendix: Hamiltonian formulation of the Dirac
equation}
\label{Dir}

Consider the Dirac equation
\begin{equation}
(i\gamma^a \nabla_a - m)\psi=0\label{Dirac}
\end{equation}
where $\nabla_a=\partial_a+ie\Phi_a$ and $\Phi_a$ is an
electromagnetic potential world-vector.
We are going to rewrite the equation in a form of the
``proper-time" covariant Hamilton equations of motion. The
``proper time" will be defined in terms of spacelike
hyperplanes constructed as follows.
 Let $\sigma_\tau(x(\tau))=0$ be an equation defining a
family of
spacelike hyperplanes. The field of timelike, future-pointing,
normalized
vectors $n^a_\tau(x)\propto\partial^a\sigma_\tau(x)$, satisfying the
continuity equation $\partial_a n^a_\tau(x)=0$, defines the
field of ``proper time" directions. Integral curves $\tau\mapsto
x^a(\tau)$ of $n^a_\tau(x)$, where $\tau$ is the parameter of the family
$\{\sigma_\tau\}$, play the role of the world-lines.
We shall need the continuity equation to guarantee the reality
of the Hamiltonian function.
Notice that this
condition eliminates some physically meaningful hyperplanes, like
the proper-time hyperboloid $\sigma_\tau(x)=x^a x_a-\tau^2=0$,
but admits simultaneity hyperplanes $\sigma_\tau(x)=n^a
x_a-\tau=0$. The ``proper time" following from the construction
should not, for this reason, be identified with the ordinary
proper time of the electron.
The ``proper time" derivative at $x$ is defined as
\begin{equation}
\frac{d}{d\tau}=n^a_\tau(x)\partial_a.
\end{equation}
Multiplying (\ref{Dirac}) from left by the Dirac matrices we
obtain \cite{Hacyan}
\begin{eqnarray}
\bigl( i\nabla^a +\sigma^{ab}\nabla_b-m\gamma^a\bigr)\psi=0.
\end{eqnarray}
Writing the four-potential explicitly in
\begin{equation}
i\partial^a\psi=\bigl(-\sigma^{ab}\partial_b +e\Phi^a
+ie\sigma^{ab}\Phi_b +m\gamma^a\bigr)\psi\label{newDirac}
\end{equation}
and contracting with $n^a_\tau(x)$ we get
\begin{eqnarray}
i\frac{d}{d\tau}\psi(x)&=&\Bigl(-\sigma_{ab}n^a_\tau(x)\partial^b
+en^a_\tau(x)\Phi_a(x)
+ie\sigma_{ab}n^a_\tau(x)\Phi^b(x)\nonumber\\
&\phantom{=}& \phantom{\Bigl(} +\,
m\,n^a_\tau(x)\gamma_a\Bigr)\psi(x)\label{newDirac'}\\
&=&\hat H\psi(x).
\end{eqnarray}
In the spinor language
\begin{eqnarray}
i\nabla^{AA'}\phi_A&=&\mu \chi^{A'}=ig_a^{\phantom{A} AA'}\nabla^a\phi_A\\
i\nabla_{AA'}\chi^{A'}&=&-\mu
\phi_{A}=ig_{aAA'}\nabla^a\chi^{A'}
\end{eqnarray}
where $\mu=m/\sqrt{2}$ and $g_a^{\phantom{A} AA'}$ are the Infeld-van
der Waerden symbols \cite{PR}.
Using the identities
\begin{eqnarray}
g^a_{\phantom{A} XA'}g^{bYA'}+g^b_{\phantom{A}
XA'}g^{aYA'}&=&g^{ab}\varepsilon _X^{\phantom{A}
Y}\\
g^a_{\phantom{A} XA'}g^{bYA'}-g^b_{\phantom{A}
XA'}g^{aYA'}&=&4\sigma^{ab\phantom{A}
Y}_{\phantom{{aa}}X}\\
g^a_{\phantom{A} AX'}g^{bAY'}-g^b_{\phantom{A} AX'}g^{aAY'}&=&4\bar
\sigma^{ab\phantom{A}
Y'}_{\phantom{{aa}}X'}
\end{eqnarray}
where $\sigma^{ab\phantom{A} Y}_{\phantom{{aa}}X}$ and $\bar
\sigma^{ab\phantom{A} Y'}_{\phantom{aa}X'}$
are generators of $(\frac{1}{2},0)$ and $(0,\frac{1}{2})$ representations of
$SL(2,{\bf C})$
 we obtain
\begin{equation}
i\nabla_a\phi_X = -4i\sigma_{abX}^{\phantom{abX}Y}\nabla^b
\phi_Y +2\mu g_{aXX'}\chi^{X'}
\end{equation}
\begin{equation}
i\nabla_a\chi^{X'} = \phantom{-}4i\bar
\sigma_{abY'}^{\phantom{abX}X'}\nabla^b
\chi^{Y'} -2\mu g_{a}^{\phantom{a} XX'}\phi_{X}
\end{equation}
and the equations obtained by their complex conjugation. These
equations are especially simple if we express generators and
Infeld-van der Waerden symbols in purely spinorial terms.
Remembering that $n^a_\tau \,n_{a\tau}=1$ implies $n_{AA'\tau}
\,n^{BA'}_{\tau}=\frac{1}{2} \varepsilon _A^{\phantom{A} B}$ we get after some
calculations
\begin{eqnarray}
i\frac{d}{d\tau}\phi_X(x)&=&i\,
n_\tau^{YY'}(x)\nabla_{XY'}\phi_Y(x)+\mu\,
n_{\tau XX'}(x)\chi^{X'}(x)\nonumber\\
&\phantom{=}& +\,e\,n_\tau^a(x)\Phi_a(x)\phi_X(x)
\\
i\frac{d}{d\tau}\chi^{X'}(x)&=&i\,
n_{\tau YY'}(x)\nabla^{YX'}\chi^{Y'}(x)-\mu\,
n_\tau^{XX'}(x)\phi_{X}(x)\nonumber\\
&\phantom{=}& +\,e\,n_\tau^a(x)\Phi_a(x) \chi^{X'}(x)
\end{eqnarray}
Let $d\sigma_\tau(x)$ be some invariant measure on
the hyperplane
$\sigma_\tau$. The equations can be derived from the Hamiltonian function
\begin{eqnarray}
H[\psi, \psi^*]&=&\langle\psi| \hat H |\psi\rangle\nonumber\\
&=&\int_{\sigma_\tau}\Bigl\{i\, \phi^*_{X'}(x)
n_\tau^{XX'}(x)n_\tau^{YY'}(x)\nabla_{XY'}\phi_{Y}(x)\nonumber\\
&\phantom{=}&\phantom{\int_{\sigma_\tau}\Bigl(}-
i\,\bar \chi^{*X}(x)
n_{\tau XX'}(x)n_{\tau YY'}(x)\nabla^{YX'}\chi^{Y'}(x)\\
&\phantom{=}&\phantom{\int_{\sigma_\tau}\Bigl(}+
\frac{1}{2}\mu\,\Bigl( \phi^*_{X'}(x)\chi^{X'}(x)+
 \chi^{*X}(x)\phi_{X}(x)\Bigr)\nonumber\\
&\phantom{=}&\phantom{\int_{\sigma_\tau}\Bigl(}+
e\,n^a_\tau(x)\Phi_a(x)n_\tau^{XX'}(x)\Bigl(\phi_{X}(x)
\phi^*_{X'}(x)+
\chi^*_{X}(x)\chi_{X'}(x)\Bigr)\Bigr\}d\sigma_\tau(x)\nonumber
\end{eqnarray}
provided
\begin{equation}
\partial^{YX'}n_{\tau XX'}(x)=\partial^{XX'}n_{\tau XY'}(x)=0\label{continuity}
\end{equation}
and the wave
functions vanish at boundaries of the hyperplane $\sigma_\tau$.
Reality of $H$ is guaranteed by the same conditions. The
Hamiltonian function is not positive definite, which is correct
since we are working here in  first quantized formalism.
 Contraction
of (\ref{continuity}) over the remaining indices implies the
continuity equation discussed above.

The explicit form of the Hamilton equations is
\begin{eqnarray}
i\,n_\tau^{XX'}(x)\frac{d}{d\tau}\phi_X(x)&=&
\frac{\delta H}{\delta \phi^*_{X'}(x)},\\
i\,n_{\tau XX'}(x)\frac{d}{d\tau}\chi^{X'}(x)&=&
\frac{\delta H}{\delta \chi^{*X}(x)},
\end{eqnarray}
and c.c, or, in the Poissonian way,
\begin{eqnarray}
i\,\frac{d}{d\tau}\phi_X(x)&=&
2\,n_{\tau XX'}(x)\frac{\delta H}{\delta \phi^*_{X'}(x)},\\
i\,\frac{d}{d\tau}\chi^{X'}(x)&=&
2\,n_\tau^{XX'}(x)\frac{\delta H}{\delta \chi^{*X}(x)},
\end{eqnarray}
We can see that $i\, n_{\tau XX'}(x)=\omega_{\tau XX'}(x)$ are  the
components of the symplectic (since derivable from a K\"ahler
potential $\parallel \psi\parallel^2$)  form on $\sigma_\tau$ at point
$x\in \sigma_\tau$, and the Poissonian form $I_{\tau
XX'}(x)=-2i\,n_{\tau XX'}(x)$.

Let $\gamma_a^{\alpha\beta}$, $a=0,1,2,3$, be the Dirac
matrices.  The Hamilton
equations equivalent to the Dirac equation written in the
bispinor form are
\begin{equation}
i\,n_\tau^{a}(x)\gamma_a^{\alpha\beta}\frac{d}{d\tau}\psi_\beta(x)=
\frac{\delta H}{\delta \psi^*_{\alpha}(x)},
\end{equation}
and c.c., where $*$ denotes the complex conjugation. The
formulas are simplest if we take simultaneity hyperplanes
foliation of the Minkowski space. Then
$n_\tau^{a}(x)\gamma_a^{\alpha\beta}=\gamma_0^{\alpha\beta}$.
Denoting its inverse by $\gamma_{0\,\alpha\beta}$ we find that
\begin{eqnarray}
\frac{d}{d\tau}\psi_\alpha(x)&=&
-i\gamma_{0\,\alpha\beta}\frac{\delta H}{\delta
\psi^*_{\beta}(x)}\\
\frac{d}{d\tau}\psi^*_\alpha(x)&=&
i\gamma_{0\,\alpha\beta}\frac{\delta H}{\delta
\psi_{\beta}(x)}
\end{eqnarray}
So here the Poissonian form is
$I_{\alpha\beta}=-i\gamma_{0\,\alpha\beta}$ and the Dirac matrix
$\gamma_{0\,\alpha\beta}$ corresponds to $\delta_{AB'}$
discussed in \ref{sec:Poisson}. The
transition to the triple bracket formalism is now
straightforward.

\end{document}